\documentclass{webofc}
\usepackage{graphicx} 
\usepackage[varg]{txfonts}
\usepackage[utf8]{inputenc}

\usepackage[frozencache,cachedir=.]{minted}

\usepackage{natbib}
\usepackage{hyperref}
\usepackage{enumitem}

\usepackage{lineno}

\newcommand{\akt}{anti-${k}_\text{T}$}
\newcommand{\Akt}{Anti-${k}_\text{T}$}

\title{Polyglot Jet Finding}

\author{\firstname{Graeme Andrew} \lastname{Stewart}\inst{1}\fnsep\thanks{\email{graeme.andrew.stewart@cern.ch}} \and
        \firstname{Philippe} \lastname{Gras}\inst{2}\fnsep \and
        \firstname{Benedikt} \lastname{Hegner}\inst{1}\fnsep \and
        \firstname{Atell} \lastname{Krasnopolski}\inst{3}
}

\institute{CERN, Esplanade des Particules 1, Geneva, Switzerland
\and
           IRFU, CEA, Université Paris-Saclay, Gif-sur-Yvette, France
\and
           Taras Shevchenko National University of Kyiv, Ukraine
          }

\abstract{%
  The evaluation of new computing languages for a large community, like HEP,
involves comparison of many aspects of the languages' behaviour, ecosystem and
interactions with other languages. In this paper we compare a number of
languages using a common, yet non-trivial, HEP algorithm: the \akt\
clustering algorithm used for jet finding. We compare specifically the algorithm
implemented in Python (pure Python and accelerated with numpy and numba), and
Julia, with respect to the reference implementation in C++, from Fastjet. As
well as the speed of the implementation we describe the ergonomics of the
language for the coder, as well as the efforts required to achieve the best
performance, which can directly impact on code readability and sustainability. }

\begin{document}

\maketitle

\section{Introduction}
\label{sec:introduction}

High energy physics (HEP), as a discipline, has undergone at least two major
shifts in language after the widespread adoption of Fortran in the
1960s~\cite{pivarski2022}. A first was a significant shift from Fortran to C++,
starting with the BaBar experiment, then gathering pace at the end of the the
Large Electron-Positron Collider (LEP) era, c.\ 2000, when the Large Hadron
Collider (LHC) experiments adopted C++ more or less wholesale. The second shift
happened with the gradual incorporation of Python into the language ecosystem of
HEP, from about 2010.

In the first transition, Fortran was almost completely displaced by C++ in the
HEP experiments; in the theory domain the evolution was more gradual and mixed, with Fortran
and C++ still both used today. In the second, a different type of transition
took place, where Python became more and more popular, but co-exists with C++.
The C++ is largely used in performance critical areas, with Python finding
traction when flexibility and rapid turn-around is needed, e.g., in
configuration and steering. Python code is typically used to interface to higher
performance C and C++ libraries, both generic (e.g., numpy) and specific HEP
codes.

Although the field is, for reasons of stability and legacy, slow to move to new
languages, there are some significant issues with the current language choices
that make an exploration of alternatives worthwhile. For example, the interfaces
between Python and C++ are a source of friction, both for passing data and error
messages back and forth, as well as being obliged to switch languages and
reimplement code on occasion, when moving from a prototype to production
(assuming the the developer actually has skills in both languages, which is not
a given). This \emph{two language problem} has potentially been addressed in the
\emph{Julia} programming
language~\cite{bib:julia_freshapproach,10.1145/3276490}, with promising
prospects for HEP, in particular, \cite{Stanitzki:2020bnx,eschle2023potential}
as well as other STEM\footnote{Science, technology, engineering, and
mathematics} areas~\cite{perkel-julia-science}. Julia offers just in time
compilation giving an ergonomic experience much like Python, but with runtime
speeds comparable to C and C++. C++ is also a notoriously tricky language to
use, particularly related to memory handling~\cite{ms-security-2019} and HEP C++
codes are frequently riddled with code defects~\cite{Naumann_2014}.

Evaluation of the prospects for a language in any particular domain area should
be done with a real problem from that domain, rather than any synthetic
benchmark. In this paper we look at the problem of jet finding, or clustering,
which is a use case from high energy physics used in calorimeter reconstruction.
This is a good example as it is not trivial, but it is also not so complex that
different implementations take too long to write.

The languages we examine here, along with links to the code used, are given in
Table \ref{tab:versions}.

\begin{table}
  \begin{center}
    \begin{tabular}{l|l}
      \textbf{Language} & \textbf{Repository} \\
      \hline
      C++ (FastJet) & \href{https://fastjet.fr/}{FastJet Website} (release 3.4.1) \\
      Python (Pure) & \href{https://github.com/graeme-a-stewart/antikt-python}{GitHub antikt-python} \\
      Python (Accelerated) & \href{https://github.com/graeme-a-stewart/antikt-python}{GitHub antikt-python} \\
      Julia & \href{https://github.com/JuliaHEP/JetReconstruction.jl}{GitHub JetReconstruction.jl} \\
    \end{tabular}
    \caption{Code repositories used in this paper. See \cite{polyglot-jets-zenodo} for exact commits and instructions.}
    \label{tab:versions}
  \end{center}
\end{table}

The evaluation itself can cover many aspects of a programming language and the
experience of using it. Metrics such as runtime are easy to evaluate, but the
ergonomics of using particular languages and the support offered by the language
ecosystem for developers are also critical and we comment on these.

\section{\Akt\ Jet Clustering Algorithm}
\label{sec:antikt}

\subsection{Algorithm}
\label{sec:alg}

The \akt\ clustering algorithm~\cite{Cacciari:2005hq,Matteo_Cacciari_2008}
is an infrared and colinear safe jet clustering algorithm, which is robust
against soft fragmentation components. We use the Fastjet
implementation~\cite{Cacciari:2011ma}, that proceeds in the following way:

\begin{enumerate}[itemsep=2pt,parsep=2pt,partopsep=0pt]
  \item A radius parameter, $R$, is defined (0.4 is typical at the LHC).
  \item For each active pseudojet $A$ (that is, an initial particle or a merged cluster):
  \begin{enumerate}
    \item Considering all other PseudoJets, $B$, which are closer in geometric
    distance than $R$, measure the minimum geometric distance:
    \begin{center}
      $d=\mathrm{min}\left( \sqrt{\Delta\eta_{AB}^2 + \Delta\phi_{AB}^2} \right)$,  
    \end{center}
    where $\Delta\eta_{AB}$ and $\Delta\phi_{AB}$ are the rapidity and azimuthal
    angle differences between $A$ and $B$.\\If there are no other pseudojets within $R$,
    then $d=R$ for pseudojet $A$.
    \item Define the \akt\ distance, $d_{ij}$, as $d_{ij} = d \,
    \mathrm{min}(k^{-2}_{\text{T},A}, k^{-2}_{\text{T},B})$ where
    $k_{\text{T},\{A,B\}}$ is the transverse momentum of the pseudojet $\{A,B\}$.
    If there is no neighbouring pseudojet, $d_{ij} = d \, k^{-2}_{\text{T},A}$.
  \end{enumerate}
  \item Choose the pseudojet with the lowest $d_{ij}$.
  \begin{enumerate}
    \item If this pseudojet has an active partner, $B$, merge these two
    pseudojets to a new pseudojet.
    \item If not, this jet is finalised and removed from the active list.
  \end{enumerate}
  \item Repeat until no pseudojets remain active.
\end{enumerate}

Note that the definition of $d_{ij}$, so-called \akt, with a negative power
favours merging jets with a high transverse momentum first, which provides stability against
soft radiation, hence its popularity. (Considering a general metric distance of
$k^{2p}_\text{T}$, $p=-1$ is \akt\ merging, $p=0$ is Cambridge/Achen merging and
$p=1$ is inclusive $k_\text{T}$, \cite{Matteo_Cacciari_2008}.)

The algorithm itself has a nice mixture of parallelisation opportunities
(pairwise matching of pseudojet candidates) and serial steps (finding the
minimum values of $d$ or $d_{ij}$), which is a good test of a non-naive
algorithm's performance.

\subsection{Algorithm Implementations}
\label{sec:algimp}

We consider two different implementations of the algorithm described above,
taken from FastJet~\cite{Cacciari:2011ma,fastjet-web}. 

The first is a \emph{plain implementation} in which, at each step, all jets are
considered as possible neighbours of each other. This algorithm has scaling that
runs roughly as $N^2$, where $N$ is the number of initial particle hits (this is
an improvement over the most naive scaling which would be
$N^3$~\cite{Cacciari:2005hq}). This implementation is fastest for $\lesssim 30$
particles.

\begin{figure}[h]
  \begin{center}
    \includegraphics[width=0.4\linewidth]{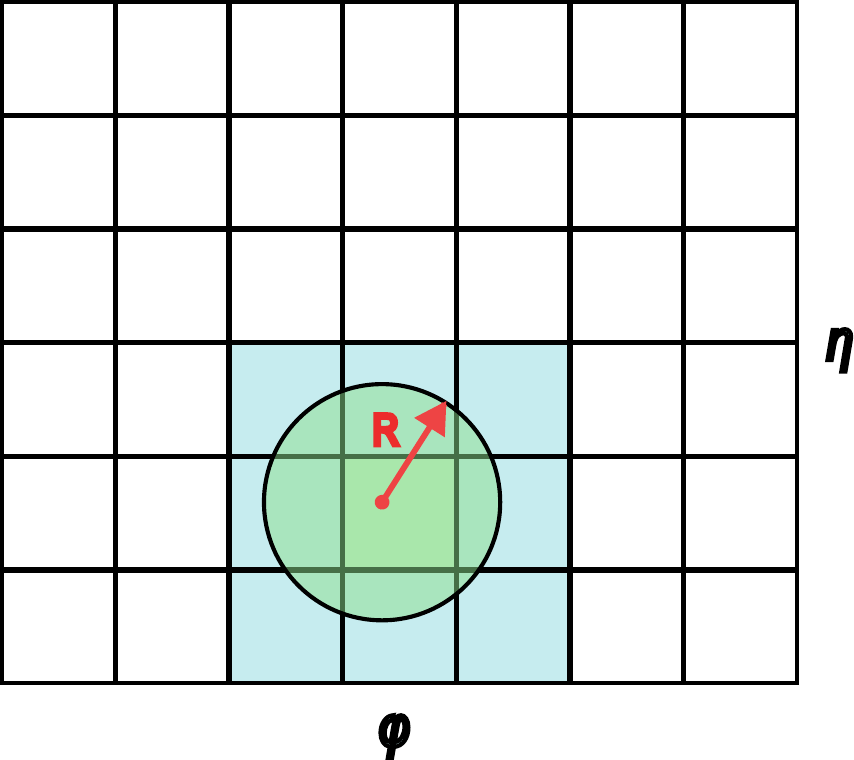}
    \caption{In the tiled algorithm implementation $(\eta,\phi)$ space is split into tiles of size $R$. When a pseudojet needs to rescan for neighbours (red dot) only pseudojets in tiles within the distance $R$ need to be considered, here shaded in light blue.}
    \label{fig:tiledimp}
  \end{center}
\end{figure}

The second is a \emph{tiled implementation}, in which the geometric space
$(\eta, \phi)$ is split into tiles of size $R$. In this way the number of
possible neighbours of any particular jet is limited to the jet's tile and to
its immediate neighbours, as illustrated in Figure \ref{fig:tiledimp}. This
strategy reduces the amount of work that needs to be done, at the expense of
extra bookkeeping of which jets are in each tile. This implementation is fastest
for p-p collisions at the LHC.

\section{Code Implementation Ergonomics}
\label{sec:ergonomics}

The code versions used are linked to in Table \ref{tab:versions}. We highlight
some specific observations in this section.

\subsection{C++, FastJet}
\label{sec:cpp-ergonomics}

The FastJet package~\cite{Cacciari:2011ma,fastjet-web} is a well maintained code
which is widely used in the HEP community. It is code which is of high quality
and well scrutinised and tested. The general style of the code is more akin to C
than C++, for reasons of minimising abstraction and increasing speed, although
templates are used extensively (where any errors are not usually nicely handled
by the compiler).

For the tiled implementation, a linked list structure is used, which requires
pointers to pointers that are challenging to reason about for the programmer, as
illustrated below.

\begin{minted}{c++}
// set up the initial nearest neighbour information
vector<Tile>::const_iterator tile;
for (tile = _tiles.begin(); tile != _tiles.end(); tile++) {
  // first do it on this tile
  for (jetA = tile->head; jetA != NULL; jetA = jetA->next) {
    for (jetB = tile->head; jetB != jetA; jetB = jetB->next) {
      double dist = _bj_dist(jetA,jetB);
      if (dist < jetA->NN_dist) {jetA->NN_dist = dist; 
        jetA->NN = jetB;}
      if (dist < jetB->NN_dist) {jetB->NN_dist = dist; 
        jetB->NN = jetA;}
    }
  }
  // then do it for RH tiles
  for (Tile ** RTile = tile->RH_tiles; RTile != tile->end_tiles; RTile++) {
    for (jetA = tile->head; jetA != NULL; jetA = jetA->next) {
      for (jetB = (*RTile)->head; jetB != NULL; jetB = jetB->next) {
        double dist = _bj_dist(jetA,jetB);
        if (dist < jetA->NN_dist) {jetA->NN_dist = dist; 
          jetA->NN = jetB;}
        if (dist < jetB->NN_dist) {jetB->NN_dist = dist; 
          jetB->NN = jetA;}
      }
    }
  }
}
\end{minted}

\subsection{Python}
\label{sec:python-ergonomics}

\subsubsection{Pure Python}

Python is renowned for being a high productivity language and implementation of
the jet finding algorithms is rather straightforward, with a clear logic.
Mutability of classes allows code to be shared between the different
implementations. e.g., an update scan for the basic case looks like this:

\begin{minted}{python3}
  def scan_for_my_nearest_neighbours(jetA: PseudoJet, jets: list[PseudoJet], 
    R2: float):
    '''Retest all other jets against the target jet'''
    jetA.info.nn = None
    jetA.info.nn_dist = R2
    for ijetB, jetB in enumerate(jets):
        if not jetB.info.active:
            continue
        if ijetB == jetA.info.id:
            continue
        dist = geometric_distance(jetA, jetB)
        if dist < jetA.info.nn_dist:
            jetA.info.nn_dist = dist
            jetA.info.nn = ijetB
    jetA.info.akt_dist = antikt_distance(jetA, jets[jetA.info.nn] 
      if jetA.info.nn else None)
\end{minted}

Where specifically \texttt{info} is a mix-in class for bookkeeping pseudojets.

While the code for the tiled implementation involves more bookkeeping, it also remains clear.

\subsubsection{Acelerated Python}

Accelerated Python code, where both numba and numpy are employed brings some
added difficulty. Not all operations are easily expressed as numpy array
calculations, particularly for dynamic arrays holding active and inactive jets.
This necessitated the use of masks, which need to be tracked. In addition, numba
jitted functions are very picky on types that can be passed (at least without
being explicitly \emph{taught} how to deal with them), so instead of a
structure, functions are called with many individual array elements, leading to
complicated call signatures. e.g., the same function as above becomes

\begin{minted}{python3}
@njit
def scan_for_my_nearest_neighbours(ijet:int, phi: npt.ArrayLike, 
                                   rap:npt.ArrayLike, inv_pt2:npt.ArrayLike, 
                                   dist:npt.ArrayLike, akt_dist:npt.ArrayLike, 
                                   nn:npt.ArrayLike, 
                                   mask:npt.ArrayLike, R2: float):
    '''Retest all other jets against the target jet'''
    nn[ijet] = -1
    dist[ijet] = R2
    _dphi = np.pi - np.abs(np.pi - np.abs(phi - phi[ijet]))
    _drap = rap - rap[ijet]
    _dist = _dphi*_dphi + _drap*_drap
    _dist[ijet] = R2 # Avoid measuring the distance 0 to myself!
    _dist[mask] = 1e20 # Don't consider any masked jets
    iclosejet = _dist.argmin()
    dist[ijet] = _dist[iclosejet]
    if iclosejet == ijet:
        nn[ijet] = -1
        akt_dist[ijet] = dist[ijet] * inv_pt2[ijet]
    else:
        nn[ijet] = iclosejet
        akt_dist[ijet] = dist[ijet] * (inv_pt2[ijet] if inv_pt2[ijet] 
          < inv_pt2[iclosejet] else inv_pt2[iclosejet])
        # As this function is called on new PseudoJets it's possible
        # that we are now the NN of our NN
        if dist[iclosejet] > dist[ijet]:
            dist[iclosejet] = dist[ijet]
            nn[iclosejet] = ijet
            akt_dist[iclosejet] = dist[iclosejet] * (inv_pt2[ijet] 
              if inv_pt2[ijet] < inv_pt2[iclosejet] else inv_pt2[iclosejet])
\end{minted}

numba also has some surprising omissions from the numpy functions which it can
JIT, e.g., array index ravelling, that required explicit reimplementation.

\subsection{Julia}
\label{sec:julia-ergonomics}

Julia is gaining in popularity because it is a language that is easy to use.
We found numerous nice features that allow code to be clear, e.g., using the
broadcast syntax for calculations on arrays is very compact:

\begin{minted}{julia}
  kt2 = (JetReconstruction.pt.(objects) .^ 2) .^ p
\end{minted}

Here \texttt{.\^} (raise to power) operates on each member of the \texttt{pt}
value of the \texttt{objects} array.

Like the FastJet code, loops can be used without sacrificing speed, so the code checking for new nearest neighbours is

\begin{minted}{julia}
# Finds new nearest neighbour for pseudojet i 
# and cross checks distance for other pseudojets back to i
# Note that nndist, near_neighbour, eta and phi are *Vectors*
function update_nearest_neighbour_crosscheck!(nndist, near_neighbour,
  i::Int, from::Int, to::Int, eta, phi, R2)
    new_nndist = R2
    new_nn = i
    @inbounds @simd for j in from:to
        delta2 = dist(i, j, eta, phi)
        if delta2 < new_nndist
            new_nn = j
            new_nndist = delta2
        end
        if delta2 < nndist[j]
            nndist[j] = delta2
            near_neighbour[j] = i
        end
    end
    nndist[i] = new_nndist
    near_neighbour[i] = nn;
end
\end{minted}

Note there are some optimisations applied here as Julia \emph{macros}, e.g.,
\texttt{@simd}, which we discuss below. In particular, here we present
updated results from those at the conference for the tiled algorithm in
Julia from applying the \texttt{LoopVectorisation} package in a key area adding
the \texttt{@turbo} macro:

\begin{minted}{julia}
find_best(diJ, n) = begin
    best = 1
    @inbounds diJ_min = diJ[1]
    @turbo for here in 2:n # Loop vectorisation marco
        newmin = diJ[here] < diJ_min
        best = newmin ? here : best
        diJ_min = newmin ? diJ[here] : diJ_min
    end
    diJ_min, best
end
\end{minted}

\section{Code Performance}
\label{sec:performance}

The different implementations of the \akt\ algorithm were tested on the
same benchmark machine, a 64 core AMD EPYC 7302 \@ 3.00GHz with 24GB RAM,
running CentOS7. The software versions used were gcc 11.3.0, Python 3.11.4 (with
numba 0.57.1 and numpy 1.24.4) and Julia 1.9.2. More details on how to reproduce
the measurements are given in \cite{polyglot-jets-zenodo}.

Reconstruction of 100 LHC-like pp events\footnote{Hard QCD $2\rightarrow2$ processes
generated with Pythia8 at 13TeV, with a minimum transverse momentum of
$20\,\mathrm{GeV}$.} was run multiple times and the average reconstruction time
per event is given in Table \ref{tab:results}. These numbers are normalised to
the FastJet tiled algorithm performance (which is $324\,\mu \mathrm{s}$ per event
on the benchmark machine). Multiple repeats of the benchmark were done and
jitter was observed to be extremely low, $<1\%$, so is not given. In these
measurements the time to read the events (in HepMC3 format) and the JIT time for
Julia and numba is excluded.

\begin{table}[h]
  \begin{center}
    \begin{tabular}{l|cc}
      \textbf{Implementation} & \textbf{Basic Algorithm} & \textbf{Tiled Algorithm} \\
      \hline
      C++ (FastJet) & 16.4 & 1.00 \\
      Python (Pure) & 504 & 110 \\
      Python (Accelerated) & 28.5 & 113 \\
      Julia & 2.83 & 0.94 \\
    \end{tabular}
    \caption{Relative run times for the reconstruction of 100 13TeV pp events, 
    normalised to the time for FastJet's tiled algorithm. Results are stable and 
    reproducible on the benchmark machine at $<1\%$.}
    \label{tab:results}
  \end{center}
\end{table}

We observe that the benchmark C++ FastJet code, with the tiled algorithm, is one
of fastest implementations. The increase in performance for the tiled
code, over the plain one, is significant with the events we used,
confirming this is both an excellent algorithm and implementation for LHC p-p
data.

As expected, the pure Python codes run very slowly in comparison. More
surprisingly, the accelerated Python codes have quite poor performance as well.
This is due to the fact that not all parts of the algorithm can be accelerated -
bookkeeping operations still run in normal Python and become dominant in the
overall runtime. This is particularly true of the tiled algorithm, which
deliberately reduces the work to be done (which can be parallelised and
accelerated) at the cost of more bookkeeping. This significantly hurts the
accelerated implementation, which ends up slower than the basic accelerated
implementation; it is not even faster than the pure Python tiled implementation
code.

Our Julia code exceeds the performance of FastJet code. In the case of the tiled
algorithm, as noted in Section \ref{sec:julia-ergonomics}, a \emph{loop
vectorisation} optimisation was applied to the search across all $d_{ij}$ to
find the minimum value, which results in a 15\% improved runtime on x86
architectures cf.\ without this macro\footnote{On Apple's M2Pro chip, the
advantage for Julia is more significant, with the final Julia code running
$\times1.45$ faster than FastJet for the tiled implementation cases}. In the case of the basic
algorithm the Julia code uses a structure of arrays layout, which the compiler
can highly optimise; additional benefit is gained from macros like
\texttt{@simd}, which allow the compiler to apply further optimisations, gaining
an additional 5\%.

There are some comments regarding these optimisations that should be made: the
Julia compiler attempts to use SIMD instructions in any case; when using the
\texttt{@simd} macro the developer is guaranteeing iterations are safe to
reorder and to overlap, and that floating point operations can be reordered;
\texttt{@turbo} also replaces some special functions with implementations that
can be vectorised better, but may be of lower accuracy. Use of these marcos may
lead to different numerical results so must be carefully validated (in our case
we have checked that they are safe). One advantage in Julia is that these
macros, as well as \texttt{@fastmath}, can be used and validated on a
case-by-case basis (cf.\ the C++ compiler options such as \texttt{-O3} or
\texttt{-ftree-vectorize}, which are applied per compilation unit, but more than
likely are actually used globally). In addition, the JIT strategy of Julia and
Python's numba will automatically target the binary architecture of the machine
being used, avoiding portability issues that can hamper C++ compiled binaries on
different microarchitectures.

\section{Conclusions}
\label{sec:conclusions}

We have implemented the \akt\ algorithm in a number of different languages
and examined code ergonomics as well as run time performance. The benchmark C++
code from FastJet is well written, but the hardest to reason on correctness, due
to the nature of the language. Python is excellent for code logic and
flexibility, but has a very poor run time performance; accelerating with numpy
and numba unfortunately takes much of this advantage away, yet still fails to
achieve a competitive run time. Julia performs extremely well, with an excellent
`out of the box' run time. The Julia compiler is able to find significant
speed-ups and features like broadcast operators help to keep code clean and
quick. Further, applying optimisations in Julia through the use of macros is
extremely easy for the programmer to exploit and result in Julia having the best
performance of all the codes that we tested.

It should be noted that the optimisations found by the Julia compiler could also
be applied to the FastJet code to close the gap. However, the authors'
experience is that doing this in C++ is considerably more difficult.

Ergonomically, C++ is also the most difficult language to use, with no package
manager, no built in profiler, and where templates and memory management remain
tricky. The breadth of libraries in C++ is impressive, although managing
dependencies is not easy. In Python the situation is far better, albeit that the
package managers are not quite standardised (pip vs.\ conda/mamba). Profiling
and debugging when accelerated code is used in Python (which is how Python is
used in data intensive science) is not easy, but package support in Python is
really excellent. In Julia the ecosystem is very well integrated, with a built
in package manager and excellent reproducibility. Julia libraries are not as
extensive as for C++ and Python, although the speed of development of new
scientific libraries (which is Julia's target community) is picking up quickly
and most areas are covered (see the discussion in Eschel et
al.~\cite{eschle2023potential}). Debugging and profiling in Julia are very well
integrated.

We conclude that expanding the use of Julia in high energy
physics would be very worthwhile, given its excellent performance and ergonomics.

\sloppy
\raggedright
\bibliography{polyglot-jets}

\begin{thebibliography}{13}

\bibitem{pivarski2022}
J.~Pivarski, \emph{History and adoption of programming languages in {NHEP}} (2022), \urlstyle{tt}\url{https://indico.jlab.org/event/505/contributions/9207/}

\bibitem{bib:julia_freshapproach}
J.~Bezanson, A.~Edelman, S.~Karpinski, V.B. Shah, SIAM Review \textbf{59}, 65 (2017), \urlstyle{tt}\url{https://doi.org/10.1137/141000671}

\bibitem{10.1145/3276490}
J.~Bezanson, J.~Chen, B.~Chung, S.~Karpinski, V.B. Shah, J.~Vitek, L.~Zoubritzky, Proc. ACM Program. Lang. \textbf{2} (2018), \urlstyle{tt}\url{https://doi.org/10.1145/3276490}

\bibitem{Stanitzki:2020bnx}
M.~Stanitzki, J.~Strube, Comput. Softw. Big Sci. \textbf{5}, 10 (2021), \texttt{2003.11952}, \urlstyle{tt}\url{https://doi.org/10.1007/s41781-021-00053-3}

\bibitem{eschle2023potential}
J.~Eschle, T.~Gal, M.~Giordano, P.~Gras, B.~Hegner, L.~Heinrich, U.H. Acosta, S.~Kluth, J.~Ling, P.~Mato et~al., \emph{Potential of the {Julia} programming language for high energy physics computing} (2023), \texttt{2306.03675}, \urlstyle{tt}\url{https://doi.org/10.48550/arXiv.2306.03675}

\bibitem{perkel-julia-science}
J.M. Perkel, Nature \textbf{572}, 141 (2019), \urlstyle{tt}\url{https://doi.org/10.1038/d41586-019-02310-3}

\bibitem{ms-security-2019}
M.~Miller, \emph{Trends, challenges, and strategic shifts in the software vulnerability mitigation landscape} (2019), \urlstyle{tt}\url{https://github.com/Microsoft/MSRC-Security-Research/blob/master/presentations/2019_02_BlueHatIL/2019_01%20-%20BlueHatIL%20-%20Trends%2C%20challenge%2C%20and%20shifts%20in%20software%20vulnerability%20mitigation.pdf}

\bibitem{Naumann_2014}
A.~Naumann, Journal of Physics: Conference Series \textbf{513}, 052023 (2014), \urlstyle{tt}\url{https://dx.doi.org/10.1088/1742-6596/513/5/052023}

\bibitem{polyglot-jets-zenodo}
G.A. Stewart, P.~Gras, B.~Hegner, A.~Krasnopolski, \emph{Polyglot jet finding - {LaTeX} sources and benchmark instructions} (2023), \urlstyle{tt}\url{https://doi.org/10.5281/zenodo.8307668}

\bibitem{Cacciari:2005hq}
M.~Cacciari, G.P. Salam, Phys. Lett. B \textbf{641}, 57 (2006), \texttt{hep-ph/0512210}, \urlstyle{tt}\url{https://doi.org/10.48550/arXiv.hep-ph/0512210}

\bibitem{Matteo_Cacciari_2008}
M.~Cacciari, G.P. Salam, G.~Soyez, Journal of High Energy Physics \textbf{2008}, 063 (2008), \urlstyle{tt}\url{https://doi.org/10.1088/1126-6708/2008/04/063}

\bibitem{Cacciari:2011ma}
M.~Cacciari, G.P. Salam, G.~Soyez, Eur. Phys. J. C \textbf{72}, 1896 (2012), \texttt{1111.6097}

\bibitem{fastjet-web}
\emph{Fastjet}, \urlstyle{tt}\url{https://fastjet.fr}

\end{thebibliography}

\end{document}